\magnification=\magstep1
\rightline{PUPT-1768$|$hep-th/9802181}
\vfil
\centerline{Minimal subtraction and the Callan-Symanzik equation} 
\vfil
\centerline{J. Naud, I. Nemenman, M. Van Raamsdonk,    and 
V. Periwal}
\medskip
\centerline{Department of Physics}
\centerline{Princeton University}
\centerline{Princeton, New Jersey 08544} 
\def\dd{\hbox{d}}

\def\curt{1}
\def\zj{2}
\baselineskip=14truept
\newcount\notenumber \notenumber=1
\def\myfootnote#1{\unskip\footnote{$^{\the\notenumber}$}{#1}%
\global\advance\notenumber by 1}
\vfil 
\par\noindent The usual proof of renormalizability using the 
Callan-Symanzik equation makes explicit use of normalization 
conditions. It is shown that demanding that the renormalization group 
functions take the form required for minimal subtraction allows one 
to prove renormalizability using the Callan-Symanzik equation, 
without imposing normalization conditions.  Scalar field theory and
quantum electrodynamics are treated.
\footline={\hfil}
\vfil\eject
\footline={\hss\tenrm\folio\hss}
An elegant and compact proof of the perturbative renormalizability of 
scalar field theory can be given using the Callan-Symanzik 
equation[\curt,\zj].
This proof constructs the renormalization group functions and 
renormalized correlation functions order-by-order without ever 
encountering an infinite quantity.
The drawback in this proof is that it makes explicit use of 
normalization conditions. This makes it awkward to establish the 
existence of renormalized Ward identities in theories with nonlinear 
symmetries, whereas these identities are trivially obtained if one 
uses minimal subtraction. Of course, on general grounds, one
knows that there is a finite renormalization that takes one from the 
renormalized theory with normalization conditions to the renormalized 
theory with minimal subtraction of infinities. However, it should be 
possible to show the renormalizability with minimal subtraction 
directly, instead of via the construction of the finite 
renormalization. This is the aim of the present paper. 

\def\pd#1{{\partial\over {\partial #1}}} \def\eps{\epsilon}
\def\al{\alpha}
\def\be{\beta}
\def\de{\delta}
\def\Ga#1{\Gamma^{#1}}
We first  consider Euclidean $\phi^{4}$ theory in $4-\eps$ dimensions. 
$\Ga{n,l}(p_{i};q_{j})$
denotes the renormalized 1PI $n$-point function with momenta $p_{i},$ 
and $l$ insertions of $-{1\over 2} \phi^{2}(q_{j}).$ To be precise, 
we show that it is possible to compute, order by order, renormalized 
correlation functions which satisfy the Callan-Symanzik equation[\zj] 
\def\cs{\hbox{(1)}}
$$\eqalign{\Big[m\pd m + (\be(g)-\eps g)\pd g -{n\over 2}\eta(g) 
-l\eta_{2}(g)\Big]
&\Ga{n,l}(p_{1},\dots,p_{n};q_{1},\dots,q_{l}) \cr = 
m^{2}(2+\delta(g))&\Ga{n,l+1}(p_{1},\dots,p_{n};q_{1},\dots,q_{l},0),\cr} 
\eqno\cs$$
with $\beta,\eta,\eta_{2},\delta$ power series in the coupling $g$ alone, and 
therefore finite order by order. This form of the renormalization 
group functions ensures that the renormalization constants are 
Laurent series in $1/\eps,$ with no finite pieces, as is appropriate 
for minimal subtraction[\zj]. In this scheme, the normalization 
conditions are replaced by
\def\norm{\hbox{(2)}}
$$\eqalign{\Ga{2,0}(p=0) &= m^{2}(1+a)\cr \pd {p^{2}}\Ga{2,0}(p=0) &= 
1+b \cr
\Ga{4,0}(p_{i}=0) &= gm^{\eps}(1+c)\cr
\Ga{2,1}(p_{i}=0;q=0) &= 1+d,\cr}\eqno\norm$$ where $a(g,\eps), 
b(g,\eps), c(g,\eps),$ and $d(g,\eps),$ are power series, at least 
$O(g)$, which we shall show to be finite as $\eps \downarrow 0.$ 
These three 1PI functions are the primitively divergent vertex 
functions in this model. We let $A_{r}$ stand for any quantity $A$ 
computed up to order $g^{r}$ in the perturbative expansion.

We write eq.~\cs\ in the form
\def\ind{\hbox{(3)}}
$$\left[m\pd m -\eps g\pd g\right]\Ga{n,l}_{r+1} = 
\left(\left[{n\over 2} \eta +l\eta_{2} - \be\pd 
g\right]\Ga{n,l}\right)_{{r+1}} +
m^{2}\left((2+\de)\Ga{n,l+1}\right)_{r+1}.\eqno\ind$$ The proof 
proceeds by induction, so it is important to make explicit the $g$ 
dependence of all quantities at the lowest order. $\be(g) = 
O(g^{2}),$ $\eta(g) = O(g^{2}),$ $\eta_{2}(g) = O(g),$ and 
$\de(g)=O(g)$[\zj]. Furthermore, $b=O(g^{2}),$ $a,c,d = O(g),$ 
$\Ga{4,1}=O(g^{2})$, and $\Ga{2,2}=O(g)$, as can readily be seen from 
the lowest order diagrams. 

The induction hypothesis is that
the primitively divergent vertex functions have been rendered finite 
up to and including $O(g^{r}),$ except for $\Ga{4,0}$ which is 
assumed finite up to order $O(g^{r+1}).$ This implies that $a_{r}, 
b_{r}, c_{r}$, and $d_{r}$ are finite. Further, we assume that 
$\be_{r+1}$ is finite, as are
$\eta_{r},\eta_{2,r},\de_{r}.$

Consider eq.~\ind\ for $n=4,l=0.$ Given the induction hypothesis and 
the fact that $\Ga{4,1}$ has a skeleton expansion, $\Ga{4,1}_{r+2}$ 
is finite. Then all terms on the r.h.s. of eq.~\ind\ are finite to 
$O(g^{r+2})$ if we can show that the combination $(2\eta g - 
\be)_{r+2}$ is finite. To show this, we evaluate~\ind\ at 
$p_{i}=0=q,$ giving \def\point{\hbox{(4)}}
$$\eqalign{(2\eta g - \be)_{r+2} + \left(\eps g^{2} \pd g 
c\right)_{r+2} = C(\eps,g)_{r+2}},\eqno\point$$ where $C_{r+2}$ is 
finite, and hence can be uniquely written as $C_{r+2} = (A(g) + \eps 
B(\eps,g))_{r+2}$, where $A_{r+2}$ and $B_{r+2}$ are finite. Thus, 
there exists a unique solution of ~\point\ with $c_{r+1}$ finite, and 
$(2\eta g - \be)_{r+2}$ finite and $\eps$ independent ({\it i.e.}, 
$(2\eta g - \be)_{r+2}=A(g)_{r+2}$, $\left(g^{2}\partial_{g} 
c\right)_{r+2}=B(\eps,g)_{r+2}$). Of course, we know nothing of the 
finiteness of $\eta_{r+1}$ or of $\be_{r+2}$ separately, but we do 
not need this information to integrate eq.~\ind\ for $n=4,l=0$ to 
obtain $\Ga{4,0}$ at arbitrary momenta. Indeed, we now see that
\def\gfour{\hbox{(5)}}
$$\left[m\pd m -\eps g\pd
g\right]\Ga{4,0}_{r+2}(p_{i};g,\eps) =
gm^{\eps}f\left({p_{i}\over m},g,\eps\right),\eqno\gfour$$ for some 
finite dimensionless function $f=O(g),$ with 
$\Ga{4,0}_{r+2}(0;g,\eps) = gm^{\eps}(1+c_{r+1})$ finite.
Since
$$\left[m\pd m -\eps g\pd g\right]\int_{0}^{1} {\dd\al\over\al} 
f\left(\al{p_{i}\over m},g\al^{\eps},\eps\right) = - 
f\left({p_{i}\over m},g,\eps\right) $$
for any function $f$ regular at zero momentum and coupling, such that 
$f(0,0,\eps)=0,$ we have
\def\gofor{\hbox{(6)}}
$$ \Ga{4,0}_{r+2}(p_{i};g,\eps) = gm^{\eps}\left[1- \int_{0}^{1} 
{\dd\al\over\al} f\left(\al{p_{i}\over 
m},g\al^{\eps},\eps\right)\right] .\eqno\gofor $$ As is 
standard[\curt,\zj], this assumes that the limit $p\downarrow 0$ does 
not introduce any pathologies into the integral over $\al,$ so that 
the finiteness already proved for $c_{r+1}$ suffices to render 
eq.~\gofor\ well defined. In perturbation theory, for $m>0,$ this 
regularity at low momenta is physically reasonable. 

The next step in the proof requires showing the finiteness of 
$\Ga{2,1}$ to $O(g^{r+1}).$ Using the skeleton expansion of 
$\Ga{2,2}$ and the induction hypothesis, eq.~\ind\ for $n=2,l=1$ has 
a r.h.s. which is finite to $O(g^{r+1})$ if the combination $(\eta + 
\eta_{2})_{r+1}$ is finite. As above, we can evaluate at $p_{i} = 0 = 
q$ to give
\def\dfour{\hbox{(7)}}
$$ -\eps g\pd g d = \left(\eta + \eta_{2} - \be\pd g d\right) 
+(\eta+\eta_{2}) d + (2+\de) m^{2}\Ga{2,2}(0;0).\eqno\dfour$$ This 
can be written as:
\def\dfive{\hbox{}}
$$(\eta+\eta_{2})_{r+1} + \left(\eps g\pd g d \right)_{r+1} = 
D(\eps,g)_{r+1}\eqno\dfive$$ where $D(\eps,g)_{r+1}$ is finite, and 
thus we can conclude that we may uniquely take $d_{r+1}$ to be 
finite, and $(\eta + \eta_{2})_{r+1}$ to be finite and 
$\eps$-independent. We can integrate eq.~\ind\ for
$n=2,l=1$ since it now takes the form
$$\left[m\pd m -\eps g\pd
g\right]\Ga{2,1}_{r+1}(p_{i};q;g,\eps) = f_{1}\left({p_{i}\over 
m},{q\over m},g,\eps\right),$$ with $f_{1}$ finite to $O(g^{r+1}),$ 
given the finiteness of $\left(\eta+\eta_{2}\right)_{r+1}$ and 
$d_{r+1}.$ 

Having shown that $\Ga{2,1}$ is finite to the next order, we can now 
consider $\Ga{2,0}.$ Eq.~\ind\ for $n=2, l=0$ has a r.h.s. which is 
finite if $\eta_{r+1}$ and $\de_{r+1}$ are finite. First note that 
$$\pd {p^{2}} \Ga{2,1}\big|_{p^{2}=0} = O(g^{2}).$$ Then Eq.~\ind\ 
for $n=2,l=0,$ gives, after differentiating with respect to $p^{2}$ 
at zero momentum,
\def\bfour{\hbox{(8)}}
$$\left(\eta + \eps g \pd g b 
\right)_{r+1}=E(\eps,g)_{r+1}\eqno\bfour$$ where $E(\eps,g)_{r+1}$ is 
finite. As above, we deduce that there exists a unique solution 
of~\bfour\ with $b_{r+1}$ finite and $\eta_{r+1}$ finite and 
independent of $\eps.$
%
%
%
%
Then with the known finiteness of
$(2\eta g -\be)_{r+2},$ and $(\eta+\eta_{2})_{r+1},$ we find that 
$\be$ is finite to $O(g^{r+2}),$ with $\eta_{2}$ finite to 
$O(g^{r+1}).$

We now consider eq.~\ind\ for $n=2, l=0$ at $p=0,$ and find after 
subtracting eq.~\dfour,
\def\afour{\hbox{(9)}}
$$ 2(a-d) -\eps g\pd g (a-d) = \eta(a-d) + (\de-\eta_{2})(1+d) 
-\be\pd g (a-d) -(2+\de)\Ga{2,2}(0;0)m^{2}.\eqno\afour$$ Observe that 
there is no way to determine $\de-\eta_{2}$ independent of $a-d.$ 
This should be expected. In minimal subtraction $\de = \eta_{2}$ is 
equivalent to $Z_{m}=Z_{2},$ ($Z_{m}$ and $Z_{2}$ are the 
multiplicative renormalization constants for the mass and the $-{1 
\over 2} \phi^{2}$ insertion, respectively).
In fact, in any scheme $Z_{m}/Z_{2}$ is a finite quantity[\zj]. In 
our present approach, we only deal with finite quantities, so we can 
consistently set $\de = \eta_{2},$ thereby determining $a$ 
unambiguously (since $d$ is already known)\myfootnote{There exists 
another solution of~\afour\ for $(a-d)$ with the $O(g^{r+1})$ term 
having an essential singularity at $\eps=0.$ In our present framework 
this solution cannot be ruled out, but it prevents the induction from 
proceeding beyond tree level. Using the explicit form of dimensional 
regularization, this solution can be rejected because an essential 
singularity in $\eps$ cannot occur at any finite order in 
perturbation theory.}. It is possible to let $\de = 
\de(g,\eps)$ and impose $a=b,$ {\it i.e.} let $m$ be the actual 
physical mass.

To complete our induction, we must exhibit a finite integral 
expression for $\Ga{2,0}$ to $O(g^{r+1}).$ Having proven the 
finiteness of $\eta_{r+1}, \de_{r+1}, a_{r+1}, b_{r+1}$, eq.~\ind\ 
for $n=2, l=0$ implies that $(\Ga{2,0}(p) - m^{2}(1+a) 
-p^{2}(1+b))_{r+1}$ satisfies an equation of the form
\def\last{\hbox{(10)}}
$$\left[m\pd m -\eps g\pd
g\right]\left(\Ga{2,0}(p) - m^{2}(1+a) -p^{2}(1+b)\right)_{{r+1}} = 
m^{2} \hat f\left({p\over m};g,\eps\right),\eqno\last$$ for a finite 
function $\hat f = O(p^{4})$ for $|p|$ small. The integrated form of 
eq.~\last\ requires showing that $$\int_{0}^{1} {\dd\al\over\al^{3}} 
\hat f\left(\al {p\over m}; g\al^{\eps},\eps\right)$$
is finite, but this is obvious from the behaviour of $\hat f$ for 
$|p|$ small.

We have therefore completed the induction step, showing that the 
Callan-Symanzik equation can be used to prove the renormalizability 
of $\phi^{4}$ theory in the minimal subtraction scheme, without ever 
imposing normalization conditions. 

\def\by{3}
We turn now to an extension of this reasoning to the case of quantum
electrodynamics.  This was considered by Blaer and Young[\by], but the
existence of renormalized Ward identities seems to have been assumed 
without discussion in their work.  In our formulation, since we use 
minimal subtraction, the existence of renormalized Ward identities is 
automatic.  In the following, we show how the Ward identities 
constrain the renormalization group functions.  The remaining steps 
then follow more or less as in Ref.~\by, and are not reproduced here.  

We follow the notation of Ref.~\zj, and consider the case of massive
Euclidean 
QED, with $m$ the mass of the photon, $M$ the mass of the electron,
and $\xi$ the gauge parameter.
The complete 1PI effective action, $\Gamma,$ may be written as a sum
 $\Gamma[A,\psi,\bar\psi;e,m,M,\xi]=\Gamma_{s}+
{1\over2}\int ( m^{2}A^{2}+ \xi^{-1} (\partial\cdot A)^{2}),$ where 
$\Gamma_{s}$ satisfies the homogeneous equation 
\def\wi{\hbox{(11)}}
\def\part{\partial}
$$ \left[\part_{\mu}{\delta\over{\delta A_{\mu}}} +ieM^{\eps/2}\left(
\psi{\delta\over{\delta\psi}} 
-\bar\psi{\delta\over{\delta\bar\psi}}\right)\right]\Gamma_{s} 
=0.\eqno\wi$$
Thus $\Gamma_{s}$ is gauge-invariant.
The general form of the Callan-Symanzik equation, differentiating 
with respect to $M,$ is
\def\qed{\hbox{(12)}}
$$\eqalign{ \Big[M&\pd M + (\be -{\eps\over 2} e)\pd e -
{n\over 2}\eta_{A}  -{k\over 2}\eta_{\psi} 
-l\eta_{2}  + {\eta_{m}\over 2}m\pd m \cr
&+\alpha \xi\pd \xi\Big]
 \Ga{n,k,l}(p_{1},\dots,p_{n};r_{1},\dots,r_{k};q_{1},\dots,q_{l})   = 
M (1+\delta ) \Ga{n,k,l+1}(p_{1}, \dots,q_{l},0),\cr} 
\eqno\qed $$
relating the proper vertex with $n$ photons, $k$ electrons, and $l$ 
insertions of $\bar\psi\psi$ to the proper vertex with one additional
insertion of $\bar\psi\psi$ at zero momentum.  There are 7 independent
renormalization group functions in this equation, 
$\alpha,\beta,\eta_{A},\eta_{\psi},\eta_{m},\delta,\eta_{2},$ all functions of 
$e,\xi,m/M,$ with no $\eps$ dependence.

For integrating eq.~\qed\ in a manner consistent with eq.~\wi, it is necessary 
that the Callan-Symanzik equation be satisfied by $\Gamma_{s},$ not 
just by $\Gamma.$    This
implies $\eta_{m}=\eta_{A}$ and $\eta_{A}=-\alpha.$ 
Further, commuting eq.~\qed\  with eq.~\wi, we find 
$\eta_{A}=2\beta/e.$  Thus, we are left with 4 independent functions,
and eq.~\qed\ simplifies to
$$\eqalign{ \Big[M\pd M   - \eps  e^{2}\pd {e^{2}} +\eta_{A}
\big\{e^{2}\pd {e^{2}}+m^{2}\pd {m^{2}}-   \xi\pd \xi-{n\over 2}\big\}
  -{k\over 2}\eta_{\psi} -&l\eta_{2}     \Big]
 \Ga{n,k,l}(p_{1},\dots  ,q_{l}) \cr  = 
M (1+\delta )&\Ga{n,k,l+1}(p_{1}, \dots,q_{l},0),\cr}  $$

The form of the QED vertex functions at zero momentum
are much restricted by eq.~\wi:
\def\normqed{\hbox{(13)}}
$$\eqalign{\Ga{2,0,0}{}^{\mu\nu}(p=0) &= m^{2}\delta^{\mu\nu} \cr 
\pd {p^{2}}\Ga{ 2,0,0}{}^{\mu}_{\mu}(p=0) &= (3-\eps)(1+a) -{1\over\xi} \cr
\Ga{ 0,2,0 }(r =0) &=  M(1+c)\cr
\pd {r^{\mu}}\Ga{0,2,0}(r =0) &= i\gamma_{\mu}(1+b)   \cr
\Ga{0,2,1}(r_{i}=0;q=0) &= 1+d,\cr
\Ga{1,2,0}{}^{\mu }(p=0;r_{i}=0) &=ieM^{{\eps/2}}\gamma^{\mu}(1+b),\cr
}\eqno\normqed$$
where $a,b,c,d$ are all functions of $e,m/M,\xi,$ {\it and} $ \eps,$ finite as
$\eps\downarrow0.$  Note that the presence of a photon mass 
ensures that there are no pathologies  associated with these 
conditions, and that we may assume analyticity at zero-momentum.
The rest of the analysis follows Ref.~\by, with  changes 
appropriate to minimal subtraction as discussed above in detail for scalar field 
theory. 

\def\cf{4}
In future work, we plan to apply this method of construction of
renormalized gauge theories to non-Abelian gauge theories, using the
Curci-Ferrari action[\cf].  
\bigskip
V.P. was supported in part by NSF grant PHY96-00258.  M.V.R was 
supported in part by an NSERC PGSA Fellowship.  J.N. was supported in part 
by an NSF Graduate Research Fellowship. 
\bigskip

\centerline{References}
\bigskip
\item{\curt.} C.G. Callan, in {\it Methods in field theory}, Les 
Houches 1975, R. Balian and J. Zinn-Justin eds. (North-Holland, 
Amsterdam 1976)

\item{\zj.} J. Zinn-Justin, {\it Quantum field theory and critical 
phenomena}, Ch. 9, 10, 18 (Second Ed.) (Oxford University Press, Oxford 1993)

\item{\by.} A.S. Blaer and K. Young, Nucl. Phys. B83, 493 (1974) 

\item{\cf.} G. Curci and R. Ferrari, Nuovo Cim. 32A, 151 (1976)

\end